# Ultrathin microwave absorbers made of mu-near-zero metamaterials

Shuomin Zhong[1] and Sailing He[1,2] *

[1]*Centre for Optical and Electromagnetic Research, State Key Laboratory of Modern Optical Instrumentations, Zhejiang University, Hangzhou 310058, China*

[2]*Department of Electromagnetic Engineering, School of Electrical Engineering, Royal Institute of Technology, S-100 44 Stockholm, Sweden*

In this paper, mu-near-zero (MNZ) metamaterials are utilized to achieve an ultrathin absorber with a thickness of only about one percent of the operating wavelength. The metamaterial absorber (MA) is made of double-layered metallic spiral arrays designed to have a large purely imaginary permeability at low microwave frequencies (~ 1.7 GHz). An absorption efficiency above 90% is demonstrated at illumination angles up to 60 degrees. A polarization-insensitive MA implemented by 2D isotropic metamaterials is also studied. Our designs have great application potential as compared with the traditional heavy and thick absorbers made of natural materials working at the same frequencies.

*Corresponding author: sailing@ kth.se

Metamaterials (MMs) can realize material properties not available in nature and thus offer unprecedented measures to manipulate waves and create exotic devices, such as negative refraction [1−3], super-lens [4] and invisible cloaks [5]. A low-index metamaterial with effective permittivity and/or permeability close to zero can give abnormal wave phenomena such as tunneling [6], radiation phase tailoring [7,8], radiation enhancement [9], and total transmission or reflection [10, 11]. MMs are usually characterized with dynamic electromagnetic properties as described by complex permittivity $\varepsilon(\omega)=\varepsilon'+i\varepsilon''$ and permeability $\mu(\omega)=\mu'+i\mu''$. Most of the research works in this field focus on low loss MMs. However, in certain circumstances the loss of MMs is also very useful for practical applications. Landy et al. first proposed to utilize lossy metamaterials to make perfect electromagnetic absorbers [12]. Since then, many different designs of metamaterial absorbers (MAs) have been proposed and demonstrated from microwave to optical frequencies [13−17].

Usually MAs are designed to have equal dynamic constituent values, i.e., $\mu(\omega)=\varepsilon(\omega)$, so that they can perfectly match the impedance in free space. Low-index MMs with large imaginary permittivity and permeability (corresponding to strong wave attenuation) are very promising in designing ultrathin absorbers with the thickness not limited by the quarter-wavelength rule, which is usually required for natural material absorbers based on interference. An absorption sheet has been experimentally fabricated with a thickness of $\lambda_0/70$ (where $\lambda_0$ is the operating free-space wavelength) [17]. Recently, Jin et al. predicted perfect absorption could be achieved by some arbitrarily thin low-index MMs with small loss [18], and this offers a new strategy to make ultrathin absorbers. In this paper we propose a different way to design ultrathin MAs by using low-index MMs with high loss $\mu(\omega)$ (with small real value) and moderate $\varepsilon(\omega)$. Although such an MM is impedance-mismatched to the free space, the metal plate backing the MM makes the total impedance matched to the free space. Unlike the design previously reported in [19], our method also utilizes compact MNZ metamaterials to achieve an ultrathin absorber with an electrical thickness of only $\lambda_0/90$.

Figures 1(a) and 1(b) illustrate the investigated structure illuminated at the transverse magnetic (TM) and transverse electric (TE) wave polarizations, respectively. A slab is sandwiched between a semi-infinite layer (air) and the bottom backing metal, which blocks the transmission. The top layer and the slab are denoted by regions 0 and 1, respectively. Consider a plane wave incident from region 0 at an incident angle $\theta$. For the TM polarization in Fig. 1(a), the nonzero field components are $H_y$, $E_x$ and $E_z$. The relative permittivity and permeability of region $n$ ($n$=0, 1) are $\varepsilon_n$ and $\mu_n$, respectively. The material of region 1 is anisotropic and its permeability tensor can be described by $\overline{\overline{\mu_1}} = \mu_{1x}\hat{x}\hat{x} + \mu_{1y}\hat{y}\hat{y} + \mu_{1z}\hat{z}\hat{z}$. By applying boundary conditions at the interfaces, one can obtain the following reflection coefficients for TM polarization and TE polarization (assuming a harmonic time dependence exp(-$i\omega t$) for the EM field),

$$r_{TM} = \frac{H_0^-}{H_0^+} = \frac{\varepsilon_1 k_{0z} + i\tan(k_{1z}d)\varepsilon_0 k_{1z}}{\varepsilon_1 k_{0z} - i\tan(k_{1z}d)\varepsilon_0 k_{1z}}$$
$$r_{TE} = \frac{E_0^-}{E_0^+} = \frac{\mu_{1x} k_{0z}\tan(k_{1z}d) - i\mu_0 k_{1z}}{\mu_{1x} k_{0z}\tan(k_{1z}d) + i\mu_0 k_{1z}},$$
(1)

where $k_n$ is the wave number in layer $n$, $k_x$ is the transverse wave number, $k_{0z} = k_0\cos\theta$, $k_x = k_0\sin\theta$, $k_0^2 = (\omega/c)^2 \mu_0\varepsilon_0$, $c = 3\times10^8 m/s$ and $k_x^2 + k_{1z}^2 = (\omega/c)^2 \mu_{1y}\varepsilon_1$ for the TM case, and $k_x^2/\mu_{1z} + k_{1z}^2/\mu_{1x} = (\omega/c)^2\varepsilon_1$ for the TE case. Since the transmission is zero, the absorption can be written as $A = 1 - R = 1 - |r|^2$, and $r = 0$ is the perfect absorption condition. In the TM case, for the normal incidence case ($\theta$=0), Eq. (1) reduces to

$$r_{TM} = \frac{\mu_0 + ik_0\mu_{1y}d}{\mu_0 - ik_0\mu_{1y}d}.$$
(2)

For simplicity, assuming that region 0 is the free space ($\mu_0$=1, $\varepsilon_0$=1), then the solution for the perfect absorption condition is

$$\mu_{1y} = ic/(\omega d).$$
(3)

From the above discussions, $\mu_{1y}$ must be equal to $ic/(\omega d)$, which is inversely proportional to thickness $d$. Similarly, we can derive the perfect absorption condition in the TE polarization case at normal incidence ($\theta$=0): $\mu_{1x} = ic/(\omega d)$. Assuming the condition is satisfied ($d = \lambda_0/100$, $\varepsilon_1 = 1$, $\mu_{1y} = 15.9i$, $\mu_{1x} = 15.9i$), we can plot the angular absorption $A$ and reflection $R$ spectra according to Eq. (1) as shown in Fig. 1(c) for the TM case and Fig. 1(d) for the TE case. One sees that the absorption decreases as incidence angle $\theta$ increases for both TM and TE polarizations. This is easy to understand from Eq. (2) since $r_{TM}$ approaches -1 (meaning total reflection) when $\theta$ approaches 90°. Nevertheless, the absorptions still remain larger than 88% when the incident angle reaches about 60°. To better understand the absorption process of the model, the simulated power flow are depicted in Fig. 2(a) and (b). They clearly show that nearly no wave is reflected by the slab for both TM and TE polarizations even at 30° incidence, which indicates good absorption by the MM layer. Here we compare the present design with two absorbers implemented by zero index MMs with $\mu(\omega)=\varepsilon(\omega)=n(\omega)$ (impedance matched to the free space) mentioned in [16] (with a large loss) and [18] (with a small loss). To make fair comparisons, we choose the same thickness of $\lambda_0/100$, and the loss part of $\varepsilon_1, \mu_1$ and $n_1$ is 15.9 for the large loss case and 0.1 for the small loss one. The angular absorptions for these three absorbers are calculated according to Eq. (1) and plotted in Fig. 2(c). Perfect absorption peaks occurs at 60° incident angle for the small loss case and 40° for the large loss one (note that the thickness of the impedance matched MA would be much thicker if the absorption peak is designed to occur at 0° incident angle). Our proposed MA has a broader angular bandwidth than the small loss case and larger absorption at normal incidence than both cases. Moreover, it is much easier to practically implement MM with a single abnormal (zero) parameter. Therefore, we see that the MA with high loss $\mu(\omega)$ and moderate $\varepsilon(\omega)$ has great advantages over absorbers studied in previous works.

The key task now is to realize the required metamaterial of region 1. According to the above analysis, the MM should be electrically small enough and with large imaginary part of $\mu(\omega)$. Here we utilize a spiral metamaterial [20] as shown in Fig. 3(a). The spiral response can be approximated based on a simple *LC* resonator model $\omega_0 = 1/\sqrt{LC}$, where $\omega_0$ is the resonant angular frequency, *L* is the inductance and *C* is the capacitance of the MM. In order to reduce the unit cell size, one must increase *L* and *C* simultaneously. We add the second spiral layer on the bottom which has 180° rotational symmetry about the *z*-axis with the top spiral (see Fig. 3(b)). Here the 180° rotation is necessary to make the inductances of the two layers added in series [20] and the capacitance *C* is also enhanced due to the interlayer interaction. The two identical spirals with different orientations are spaced by a dielectric substrate made up of Rogers Printed circuit board (PCB). The parameters of the unit cell are as follows: dimensions of unit cell $p_x$ = 4.1 mm and *d* = 1.9 mm. We retrieve the effective parameters of the metamaterials slab through the transfer-matrix method approach [21] using the simulated scattering parameters, $S_{11}$ and $S_{21}$. The effective permittivity $\varepsilon_1$ keeps nearly a constant of 2.87 with small loss, which is not shown here. The extracted effective permeability is shown in Fig. 3(c). The unit cell exhibits a Lorentz-type magnetic response near the resonance frequency 1.68 GHz. Figure 3(d) shows the absorption and reflection of MA using effective Lorentz-type resonance MMs. An absorption peak occurs at 1.68GHz where the perfect absorption condition is satisfied with $\mu=0+14i$, which confirms the validity of our analytical model.

The simulations are carried out using a frequency domain solver, implemented by CST Microwave Studio TM 2010[22]. Figure 4(a) is a three-dimensional illustration of the simulated MA. In the simulation, all metallic components are modeled as copper with a conductivity of $\sigma = 5.8 \times 10^7$S/m and the dielectric constant and loss tangent of the dielectric spacer are 2.9 and 0.0025, respectively. A TM wave with the magnetic field polarized along the *y* direction is chosen as the excitation source. Periodic boundary conditions are set in the *x* and *y* directions. Since the transmission is totally blocked by the bottom copper plate, the absorption can be calculated as $A=1-R$.

The MMs are fabricated by using the standard PCB photolithography. All fabricated dimensions are identical to those simulated and the photograph is shown in Fig. 4(c). The MA has 50 unit cells in the *x* direction and 48 strips in the *y* direction with an area of 200mm×300mm. The period in the *x* and *y* directions are the same, i.e. $p_x = p_y$. The absorption performance of the tested sample in Fig. 4(c) is verified by measuring the complex *S*-parameters in a microwave anechoic chamber. The experimental setup is the same as that in [23]. Two horn antennas are connected to a R&S ZVB vector network analyzer as the transmitter and receiver. The measured reflection is normalized with respect to a copper plane with the same dimensions as the sample. We add extra absorbers on the two side regions of the sample without MMs to reduce the unwanted reflection. The same method is also adopted to the copper reference plane. As depicted in Fig. 4(d), a 94% absorption peak is observed at 1.73GHz in the experiment (red dashed curve), which agrees well with the simulation (black solid curve), considering various errors in the fabrication and assembly. The operating frequency deviation from the effective medium model is due to the influence of the metallic back, which is now very close (a bit inside the effective medium boundary of the MM) to the magnetic inclusions in order to compress the thickness of the MA.

At last, we discuss the angular and polarization properties of the MAs. Figures 5(a) and 5(b) show the angular absorption spectrums of the MA of Fig. 4(a). Four incident angles, namely, 10°, 30°, 45° and 60°, are selected. The experimental results agree well with the simulation results, which demonstrate over 90% absorption at all these incidence angles. According to the above analytical model, absorption for both TM and TE polarizations can be achieved when both $\mu_{1y}$ *and* $\mu_{1x}$ are equal to $ic/(\omega d)$. Since the period of our metamaterial slab in the *x* and *y* directions are equal, it is easy to obtain a polarization-independent absorber by packing them into a 2D isotropic unit cell as shown in Fig. 4(b). Figures 5(c) and 5(d) are the simulated results of the angular absorption of the polarization independent MA (see Fig. 4(b)) for TM and TE polarizations, respectively. The incident angle is varied from 0° to 80° in a step of 10°. Absorption peaks are observed at 1.7GHz for both polarizations. The frequency shift of the absorption peaks is tiny at different incident angles below 60° and become larger at incident angles over 60°. This originates from the anisotropy of the MM at a large oblique incident angle and numerical errors (due to limited mesh accuracy) in our simulations. Nonetheless, the numerical results show that the absorption remains over 90% when the incident angle is smaller than 60°, which confirms the analytical model in Fig.1.

In conclusion, an analytical model revealing perfect absorption by an ultrathin MNZ layer has been proposed. Then we have fabricated ultrathin metamaterial absorbers with the electrical thickness of only $\lambda_0$/90. The benefit of thin absorbers, compared with traditional absorbers made of natural materials, brings great potential in many applications. This concept can be extended to other frequencies, such as terahertz, infrared and optical frequencies.

**Acknowledgements**

We acknowledge Zuojia Wang and Prof. Hongsheng Chen from the Electromagnetic Academy at Zhejiang University for their great help in the sample test. Shuomin Zhong would like to thank Fei Ding, Dr. Yi Jin and Prof. Yungui Ma for fruitful discussions. This work is partially supported by the National High Technology Research and Development Program (863 Program) of China (No. 2012AA030402), the National Natural Science Foundation of China (Nos. 61178062 and 60990322), Swedish VR grant (# 621-2011-4620) and AOARD.

**References**

[1] J. B. Pendry, A. T. Holden, W. J. Stewart, and I. Youngs, Phys. Rev. Lett. **25**, 4773 (1996).

[2] J. B. Pendry, A. J. Holden, D. J. Robbins, and W. J. Stewart, IEEE Trans. Microwave Theory Tech. **47**, 2075 (1999).

[3] R. A. Shelby et al., Science **292**, 77 (2001).

[4] J. B. Pendry, Phys. Rev. Lett. **85**, 3966 (2000).

[5] J. B. Pendry, D. Schurig, and D. R. Smith, Science **312**, 1780, (2006).

[6] M. G. Silveirinha and N. Engheta, Phys. Rev. Lett. **97**, 157403 (2006).

[7] A. Alu`, M. G. Silveirinha, A. Salandrino, and N. Engheta, Phys. Rev. B **75**, 155410 (2007).

[8] Y. G. Ma, P. Wang, X. Chen, and C. K. Ong, Appl. Phys. Lett. **94**, 044107 (2009).

[9] Y. Jin and S. L. He, Opt. Express **18**, 16587 (2010).

[10] X. Huang, Y. Lai, Z. H. Hang, H. Zheng, and C. T. Chan, Nat. Mater. **10**,582 (2011).

[11] J. Hao, W. Yan, and M. Qiu, Appl. Phys. Lett. **96**, 101109 (2010).

[12] N. I. Landy, S. Sajuyigbe, J. J. Mock, D. R. Smith and W. J.Padilla, Phys. Rev. Lett. **100**, 207402 (2008).

[13] Y. Q. Ye, Y. Jin, and S. L. He, J. Opt. Soc. Am. B **27**, 498 (2010).

[14] X. L. Liu, T. Starr, A. F. Starr, and W. J. Padilla, Phys. Rev. Lett. **104**, 207403(2010).

[15] F. Ding, Y. X. Cui, X. C. Ge, Y. Jin, and S. L. He, Appl. Phys. Lett. **100**, 103506 (2012).

[16] C. M. Watts, X. L. Liu, and W. J. Padilla, Adv. Mater. **24**, OP98 (2012).

[17] Y. Cheng, H. Yang, Z. Cheng, N. Wu, Appl. Phys. A, **102**, 99 (2011).

[18] Y. Jin, S. S. Xiao, N. A. Mortensen, and S. L. He, Opt. Express **19**, 11114 (2011).

[19] K.B. Alici, F. Bilotti, L. Vegni, and E. Ozbay, J. Appl. Phys. **108**, 083113 (2010).

[20] W.-C. Chen, C. M. Bingham, K. M. Mak, N. W. Caira, and W. J. Padilla, Phys. Rev. B **85**, 201104(R) (2012).

[21] X. D. Chen, T. M. Grzegorczyk, B. I. Wu, J. Pacheco, Jr. and J. A. Kong, Phys. Rev. E **70**, 016608 (2004).

[22] CST Studio Suite 2010, CST of America, 2010.

[23] L. Huang and H. S. Chen, Progress In Electromagnetics Research **113**, 103 (2011).

**Captions**

Figure 1
(Color online). Configuration of the theoretical models for (a) TM polarization and (b) TE polarization. Analytical angular absorption (red line) and reflection (blue line) for (c) TM polarization and (d) TE polarization.

Figure 2
(Color online). (a) and (b) are power flow for TM and TE cases with θ=30°, respectively. (c) Angular absorption performance of three different MAs. ($\varepsilon_1, \mu_1$) has values of (1, 15.9$i$), (15.9$i$, 15.9$i$) and (0.1$i$, 0.1$i$) for black, red and blue lines, respectively.

Figure 3
(Color online). (a) Schematic illustrating the geometry of the spiral. (b) Schematic (exploded view) of the dual-layer square spiral metamaterial (top) and a side view showing material layers (bottom). (c) The effective permeability of the unit cell. The black solid line and red dashed line denote real part and imaginary part, respectively. (d) The absorption (red line) and reflection (blue line) of MA, calculated with effective Lorentz-type resonance MMs.

Figure 4
(Color online). (a) 3D illustration of the MA structure for one polarization (each unit cell is formed by the dual-layer square spiral metamaterial shown in Fig. 3(b)). (b) 3D illustration of the polarization-independent MA structure. (c) Photograph of the fabricated sample. (d) Comparison between the experimental absorption (red line) and simulated absorption (black line) of the MA in Fig. 4(a) at normal incidence.

Figure 5
(Color online). (a) Simulated angular absorption of the MA in Fig. 4(a). (b) Experimental angular absorption of the MA in Fig. 4(c). (c) Simulated angular absorption of the MA in Fig. 4(b) for TM polarization. (d) Simulated angular absorption of the MA in Fig. 4(b) for TE polarization.

Figure 1

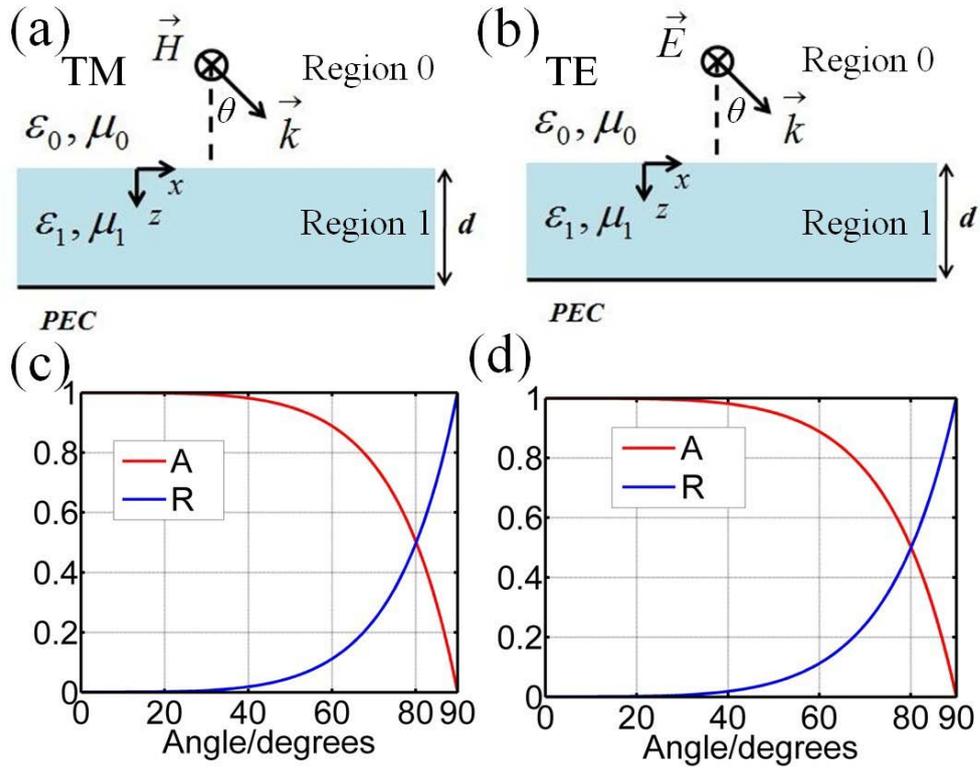

Figure 2

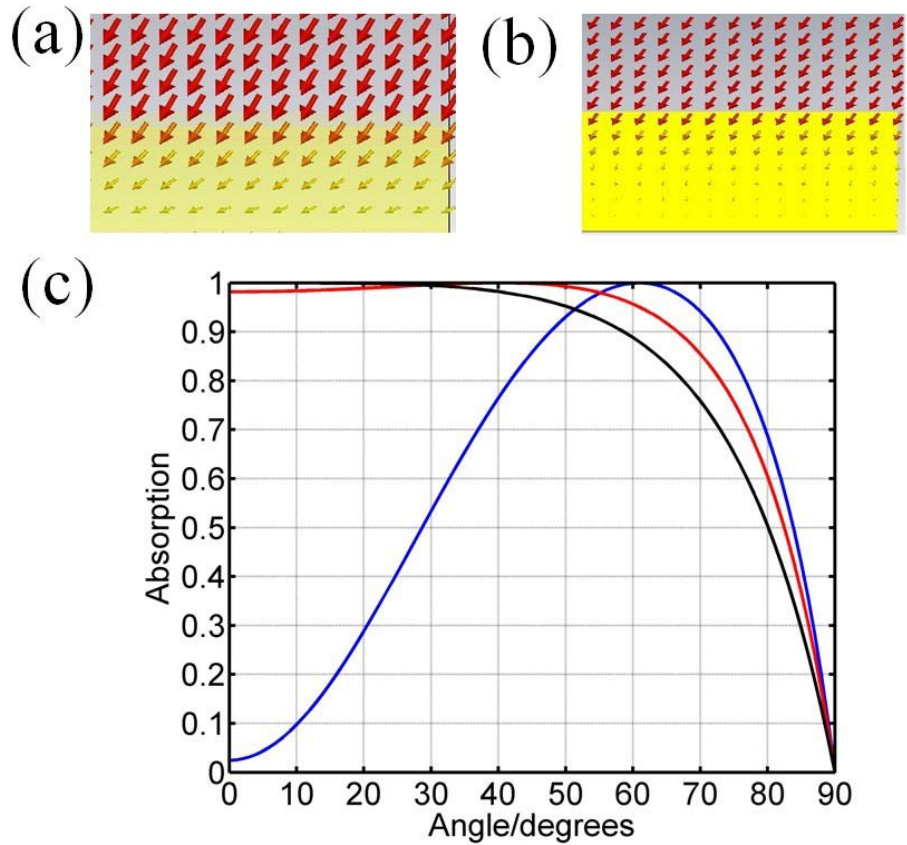

Figure 3

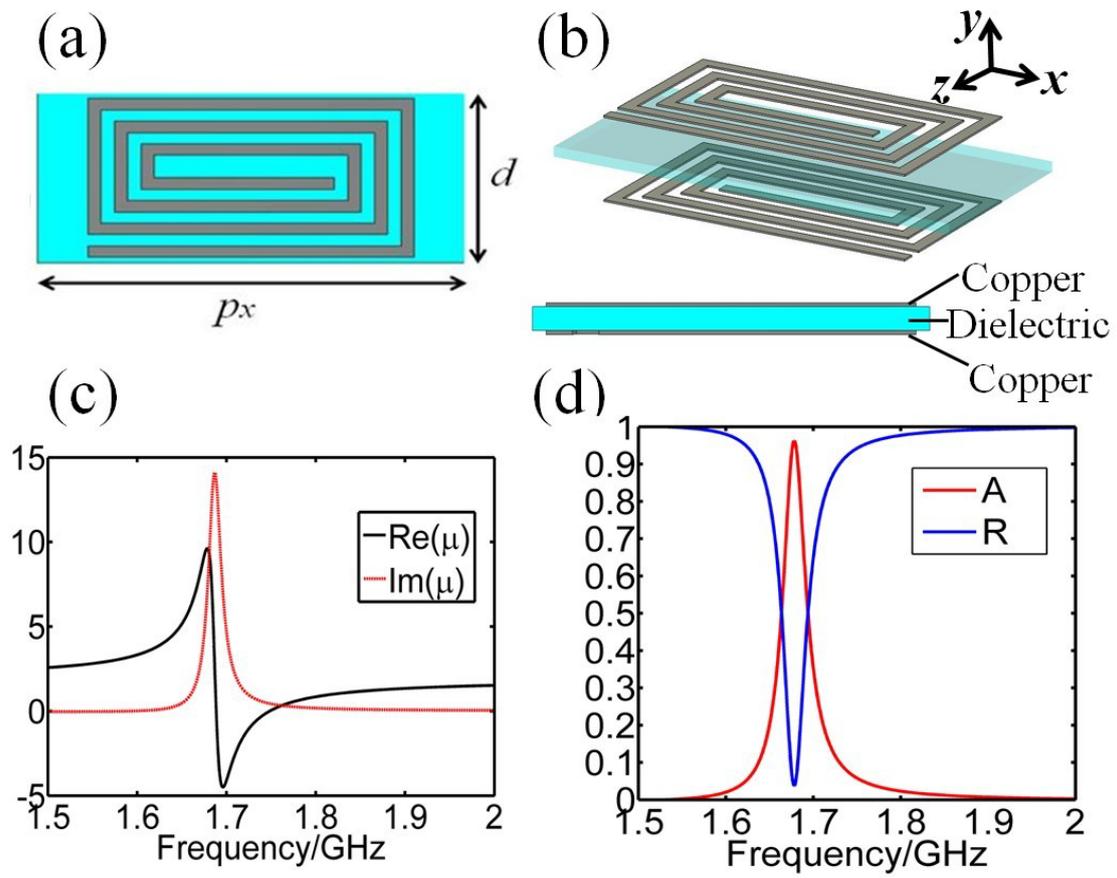

Figure 4

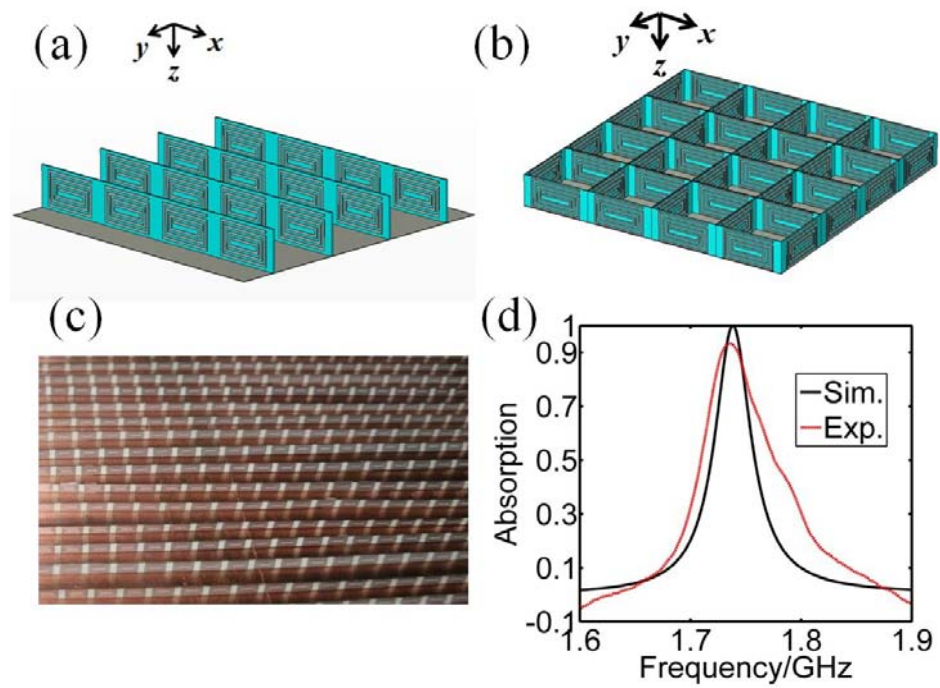

Figure 5

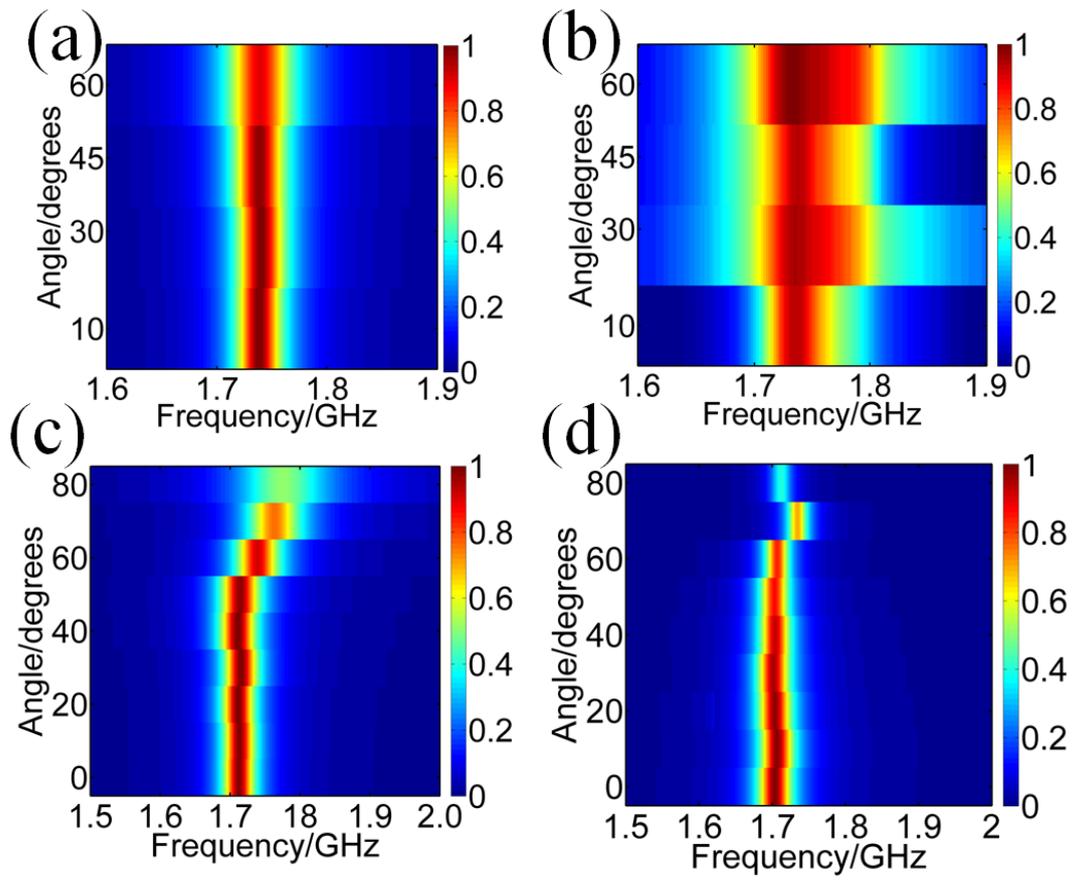

11